\documentclass[aps,prl,twocolumn,reprint,groupedaddress,showpacs]{revtex4-1}

\usepackage{amsmath}
\usepackage{amssymb}
\usepackage{graphicx}
\usepackage{color} 
\usepackage[normalem]{ulem} 
\usepackage{newfloat}

\DeclareFloatingEnvironment[name={FIG. S.\!\!}]{suppfigure}
\DeclareFloatingEnvironment[name={TABLE S.\!\!}]{supptable}
\DeclareMathAlphabet{\mathpzc}{OT1}{pzc}{m}{it}

\newcommand{\ket}[1]{\left|#1\right\rangle}

\newcommand{\ketbra}[2]{\left| #1 \right\rangle \! \left\langle #2 \right|}
\newcommand{\statAverage}[1]{\left< \! \left< #1 \right> \! \right>}

\begin{document}


\title{Soft Decoding of a Qubit Readout Apparatus}


\author{B.~D'Anjou}
\affiliation{Department of Physics, McGill University, Montreal, Quebec, H3A 2T8, Canada}
\author{W.A.~Coish}
\affiliation{Department of Physics, McGill University, Montreal, Quebec, H3A 2T8, Canada}
\affiliation{Canadian Institute for Advanced Research, Toronto, Ontario, M5G 1Z8, Canada}



\date{\today}

\begin{abstract}
Qubit readout is commonly performed by thresholding a collection of analog detector signals to obtain a sequence of single-shot bit values. The intrinsic irreversibility of the mapping from analog to digital signals discards soft information associated with an \emph{a posteriori} confidence that can be assigned to each bit value when a detector is well characterized. Accounting for soft information, we show significant improvements in enhanced state detection with the quantum repetition code as well as quantum state or parameter estimation. These advantages persist in spite of non-Gaussian features of realistic readout models, experimentally relevant small numbers of qubits, and finite encoding errors. These results show useful and achievable advantages for a wide range of current experiments on quantum state tomography, parameter estimation, and qubit readout.
\end{abstract}


\maketitle

\begin{figure}
\centering
\includegraphics[width=\columnwidth]{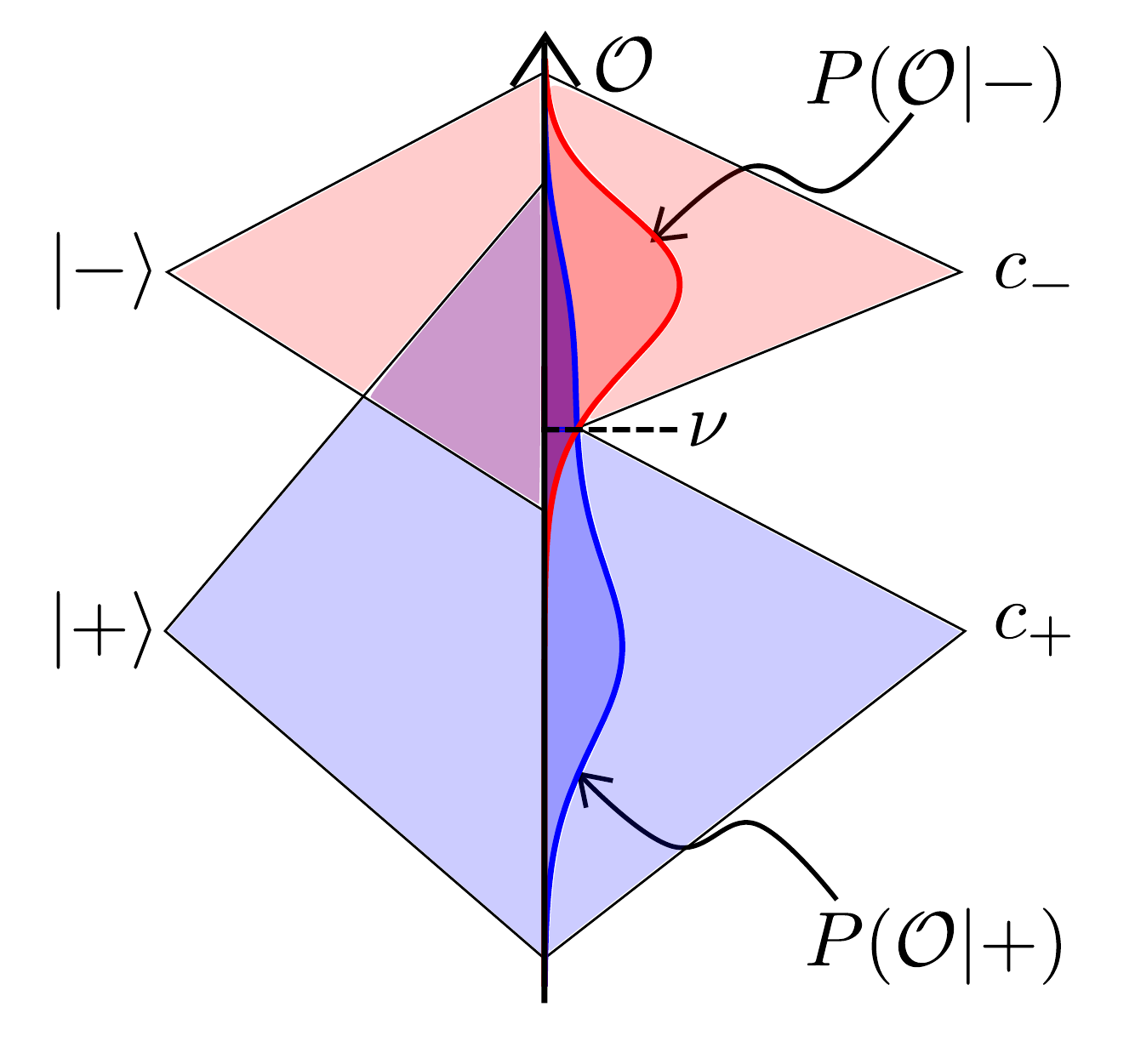}
\centering
\caption{(Color online) The single-shot readout thresholding procedure (threshold $\nu$) erases information by irreversibly converting a continuum of analog readout outcomes $\mathcal{O}$ into a single binary value $c_{\pm}$. The conditional probability distributions $P(\mathcal{O}|\pm)$ are typical of the readout performed in Refs.~\cite{elzerman2004,morello2010} and described in Ref.~\cite{danjou2014}.\label{fig:fig1}}
\end{figure}

In most quantum-measurement tasks, the goal is to extract information encoded in a stream of quantum states. For a single-qubit expectation value, the stream is a collection of identically prepared single qubits. To perform a Bell-inequality measurement, the stream is a collection of entangled qubit pairs. For quantum error detection or correction, the stream consists of many qubits that make up the code, on which multiqubit syndrome measurements are performed. In practice, in all of these scenarios, information is commonly extracted by measuring individual qubits or joint observables in a single shot~\cite{haffner2005,schindler2011,bernien2013,chow2011,reed2012,chow2013,saira2013,barends2014}. While this strategy can be optimal for extracting a single bit of information, e.g., the state $\ket{\pm}$ of a single qubit, it is generally suboptimal when considering streams of data. A single-shot qubit readout typically involves the irreversible conversion of an analog outcome $\mathcal{O}$ from a readout apparatus (e.g., a current or voltage pulse, the quadrature of a microwave tone, etc.) into a binary outcome $c_{\pm}$ via thresholding~\cite{myerson2008,barthel2009,morello2010,neumann2010,robledo2011,pla2013,lin2013,liu2014,harty2014} (see Fig.~\ref{fig:fig1}). Thresholding erases information about the posterior probability $P(\pm|\mathcal{O})$ that can be ascribed to each bit value given $\mathcal{O}$. In contrast, when a sequence of analog readout outcomes $\mathcal{O}$ is fed into a decoder that accepts analog values as input, the frequency of decoding errors can be significantly reduced~\cite{chase1972}. Such \emph{soft-decoding} techniques have been central to the development of capacity-achieving classical codes \cite{guizzo2004} now used in deep-space communications and high-bandwidth 3G/4G cellular networks. Soft-decision decoding has been applied to quantum codes~\cite{poulin2006,duclos-cianci2010} and to schemes for fault-tolerant quantum computing~\cite{goto2013}, in which the soft decision is made by correlating multiple single-shot qubit readout outcomes. Soft decoding has also been identified as an important tool for continuous-variable quantum key distribution~\cite{mondin2010}.

While soft-decoding methods are routinely applied to classical noisy signals for communication applications, they have not seen widespread application in qubit readout methods. Here, we exploit the fact that the qubit readout itself can be treated as a communication channel characterized by a pair of conditional probability distributions $P(\mathcal{O}|\pm)$ for the analog signal $\mathcal{O}$ (see Fig.~\ref{fig:fig1}), \emph{even if} the readout apparatus is designed to perform a binary measurement. Soft decoding of a readout apparatus can lead to significant improvements in a number of quantum-information tasks. To achieve these improvements, the physical characteristics of the readout dynamics and the noise must be well understood to determine the distributions $P(\mathcal{O}|\pm)$. Readout errors are especially sensitive to the tails of these distributions, so it is important to understand non-Gaussian features (fat tails or bimodality) to reap the benefits of soft decoding. Crucially, the distributions $P(\mathcal{O}|\pm)$, which must already be known to characterize the single-shot readout fidelity, can be measured or modeled accurately for several state-of-the-art qubit implementations~\cite{myerson2008,barthel2009,harty2014,neumann2010,robledo2011,morello2010,pla2013,veldhorst2014,lin2013,liu2014,jeffrey2014,gambetta2007,danjou2014}.

In this Letter, we explicitly demonstrate the advantages of soft decoding through two experimentally significant examples: enhanced state detection via the quantum repetition code, and state or parameter estimation. In particular, the number of qubits required for efficient enhanced state detection can be reduced, through data processing alone, by up to a factor of $2$, an advantage which persists for a small number of qubits and finite encoding errors. Because additional qubits are an expensive resource, this result is of immediate practical importance. We also extend a result of Ref.~\cite{ryan2013} by making use of soft information to further improve the precision of measurements of Pauli operators required for state tomography. In both cases, we benchmark the improvement by comparing the performance of the widely used and efficient maximum-likelihood estimation~\cite{cramer1946} when applied to analog instead of thresholded qubit readout outcomes. Crucially, we find and characterize significant improvements not only for the idealized Gaussian readout, but also for the realistic non-Gaussian readout investigated in Ref.~\cite{danjou2014} and relevant to many experiments~\cite{elzerman2004,morello2010,pla2013,veldhorst2014}.

\emph{Enhanced state detection.$-$} In the quantum repetition code, a logical qubit (with basis $\left\{\ket{0},\ket{1}\right\}$) in the state $\ket{\psi}=\alpha_0 \ket{0} + \alpha_1 \ket{1}$ is encoded into $N$ physical qubits (with basis $\left\{\ket{+},\ket{-}\right\}$): $\ket{\psi_N} = \alpha_0 \ket{-}^{\otimes N} + \alpha_1 \ket{+}^{\otimes N}$~\cite{deuar1999,divincenzo2005,schaetz2005}. The redundant outcomes of independent measurements of the $N$ physical qubits are then correlated to reduce measurement errors. The simplest approach is to measure each qubit in a single shot and assign the binary value $c_i=c_{\pm}$ to the $i$th qubit. When the encoded states $\ket{0}$ and $\ket{1}$ are equally likely \emph{a priori}, the optimal approach is to calculate the likelihood ratio~\cite{wozencraft1965} of the data $\left\{c_i\right\}$:
\begin{align}
	\Lambda_c \equiv \frac{P(\left\{c_i\right\}|1)}{P(\left\{c_i\right\}|0)} = \prod_{i=1}^{N} \frac{P(c_i|+)}{P(c_i|-)}. \label{eq:thresholdLikelihoodRatio}
\end{align}
In Eq.~\eqref{eq:thresholdLikelihoodRatio}, $P(c_i|\pm)$ is the probability to obtain the value $c_i$ given that the $i$th qubit is in the state $\ket{\pm}$. The projected state of the logical qubit is most likely $\ket{1}$ ($\ket{0}$) if $\Lambda_c > 1$ ($\Lambda_c < 1$). We may rewrite the likelihood ratio, Eq.~\eqref{eq:thresholdLikelihoodRatio}, in terms of the conditional single-shot error rates $\epsilon_\pm \equiv P(c_\mp|\pm) < 1/2$:
\begin{align}
	\Lambda_c = \left(\frac{1-\epsilon_+}{\epsilon_-}\right)^{n_+}\cdot \left(\frac{\epsilon_+}{1-\epsilon_-}\right)^{N-n_+} , \label{eq:majorityVoteEstimator}
\end{align}
where $n_+$ is the number of measurements for which the outcome $c_+$ occurred. For a binary symmetric readout, $\epsilon = \epsilon_{\pm}$, Eq.~\eqref{eq:majorityVoteEstimator} results in a simple majority vote since $n_+ > N/2$ ($n_+ < N/2$) implies $\Lambda_c > 1$ ($\Lambda_c < 1$).

Equation~\eqref{eq:majorityVoteEstimator} is a maximum-likelihood estimator applied to single-shot readout outcomes. However, a physical readout apparatus typically yields an analog readout outcome $\mathcal{O}_i$ that need not be thresholded to a binary value $c_i$. The observable $\mathcal{O}_i$ could be, for example, the time average of a fluorescence signal~\cite{schaetz2005,myerson2008,robledo2011}, the peak of a current pulse through a single-electron transistor or quantum point contact~\cite{elzerman2004,morello2010,veldhorst2014}, the quadrature of a microwave tone~\cite{lin2013,liu2014,jeffrey2014}, or even the likelihood ratio of a single-shot readout~\cite{gambetta2007,hume2007,myerson2008,harty2014,danjou2014}. Thresholding leads to an irreversible loss of information about the confidence $P(\pm|\mathcal{O}_i)$ in each bit value. Soft decoding, which makes full use of that information, is achieved by instead applying the maximum-likelihood estimator to the analog readout outcomes:
\begin{align}
	\Lambda_{\mathcal{O}} \equiv \frac{P(\left\{\mathcal{O}_i\right\}|1)}{P(\left\{\mathcal{O}_i\right\}|0)} = \prod_{i=1}^{N} \frac{P(\mathcal{O}_i|+)}{P(\mathcal{O}_i|-)}, \label{eq:analogLikelihoodRatio}
\end{align}
where $P(\mathcal{O}_i|\pm)$ is the probability density for outcome $\mathcal{O}_i$ given that the $i$th qubit is in the state $\ket{\pm}$.

To take full advantage of soft decoding, it is necessary to have an accurate representation of the conditional probability distributions $P(\mathcal{O}|\pm)$ for the analog qubit readout outcomes. A common idealization for a readout is the Gaussian readout, $P(\mathcal{O}|\pm)=\sqrt{r/2\pi}\exp{\left[-(\mathcal{O}\mp 1)^2 r/2\right]}$, where $r$ is the power signal-to-noise ratio. Soft decoding of the Gaussian readout with maximum-likelihood estimators such as Eqs.~\eqref{eq:thresholdLikelihoodRatio} and \eqref{eq:analogLikelihoodRatio} has been extensively studied in the context of classical communication theory~\cite{wozencraft1965,chase1972}. Since a projective measurement collapses $\ket{\psi_N}$ to either $\ket{+}^{\otimes N}$ or $\ket{-}^{\otimes N}$, the advantage obtained by soft decoding of the readout apparatus translates directly to the quantum case. More precisely, for $r \gg 1$, the number of qubits $N_c$ and $N_\mathcal{O}$ required to achieve a target error rate $\varepsilon$ using $\Lambda_c$ and $\Lambda_\mathcal{O}$, respectively, are related (for $N_c$ odd) by \cite{supplemental}:
\begin{align}
	N_{\mathcal{O}} = \frac{N_c+1}{2} + \frac{N_c-1}{2}\frac{\ln r}{r} + O\left(\frac{N_c}{r}\right). \label{eq:asymptoticAdvantage}
\end{align}
Thus, soft decoding can reduce the number of required physical qubits by up to a factor of $2$ compared to the majority vote (asymptotically, $N_c\sim 2N_\mathcal{O}$, or alternatively, for fixed $N$, $\varepsilon_{\mathcal{O}}\sim \varepsilon_c^2$ up to a logarithmic prefactor in $\varepsilon_c$). Intuitively, this advantage arises since the majority vote ignores all information contained in strings for which more than half of the bits are corrupted, while soft decoding associates every string with some confidence. A similar asymptotic advantage exists for arbitrary block codes transmitted through a Gaussian communication channel~\cite{chase1972}. Importantly, Eq.~\eqref{eq:asymptoticAdvantage} is valid for the regime of reasonably small $N$ relevant to recent experiments~\cite{monz2011,schindler2011,reed2012,chow2013,saira2013,barends2014,goodwin2014}. Moreover, the form of the subleading corrections in Eq.~\eqref{eq:asymptoticAdvantage} suggests that they can be small for realistic experimental values of $r$. We have indeed verified, using the exact analytical expressions for the error rates \cite{supplemental}, that an advantage persists for low signal-to-noise ratio and relatively small $N$. For example, we require $N_{\mathcal{O}} = 6$ instead of $N_c = 9$ to reach an error rate $\varepsilon < 3 \times 10^{-4}$ for $r = 2$. 

Realistic qubit readouts are typically not well represented by Gaussian probability distributions~\cite{gambetta2007,myerson2008,morello2010,robledo2011,danjou2014}. To verify that soft decoding of the readout apparatus still provides an advantage in experimentally relevant cases, we apply the estimators in Eqs.~\eqref{eq:thresholdLikelihoodRatio} and \eqref{eq:analogLikelihoodRatio} to the realistic non-Gaussian ``peak-signal'' readout implemented in Refs.~\cite{elzerman2004,morello2010} and for which the distributions $P(\mathcal{O}|\pm)$ were analyzed in Ref.~\cite{danjou2014}. In this measurement, the analog outcome $\mathcal{O}$ is the peak value of a finite-duration current pulse signalling the excited state $\ket{+}$ and subject to Gaussian white noise (see Ref.~\cite{supplemental} for a summary). A typical pair of distributions for this readout is illustrated in Fig.~\ref{fig:fig1}. Even though the distribution $P(\mathcal{O}|+)$ has strong non-Gaussian features, soft decoding still gives an appreciable advantage. For example, Monte Carlo simulations with $10^6$ random records show that, similar to the Gaussian readout, we require $N_{\mathcal{O}} = 6$ instead of $N_c = 9$ to reach an error rate $\varepsilon < 0.05$ for a signal-to-noise ratio $r=2$ \cite{supplemental}.

 To account for errors during encoding, we allow for uncorrelated bit flips with probability $\eta$ for both states $\ket{\pm}$. The probability distributions for the analog readout outcomes then become $P(\mathcal{O}_i|1/0) = (1-\eta)P(\mathcal{O}_i|+/-) + \eta P(\mathcal{O}_i|-/+)$, giving a modified version of the likelihood ratio, Eq.~\eqref{eq:analogLikelihoodRatio}. We find that when encoding errors $\eta$ are sufficiently large, the soft-decoding procedure reduces to a simple thresholding procedure~\cite{supplemental}. However, for the Gaussian readout, soft decoding can still give an advantage over thresholding if
\begin{align}
	\eta \lesssim e^{-2 r}
\end{align}
when $r\gg 1$. Thus, the encoding bit-flip rate must merely be smaller than some power of the single-shot readout error rate $\epsilon$ ($\epsilon \sim e^{-\frac{r}{2}}$ up to logarithmic corrections). To verify this, we have performed a Monte Carlo simulation of the error rate for the Gaussian readout by generating $10^7$ random measurement records taking into account the bit-flip rate $\eta$ \cite{supplemental}. For example, we find that for $r = 2$ and $\eta= 1\%$, we require $N_{\mathcal{O}} = 6$ instead of $N_{c}= 9$ qubits to achieve $\varepsilon < 8 \times 10^{-4}$. Similarly, for the peak-signal readout described in Ref.~\cite{danjou2014}, we find from a simulation of $10^6$ random measurement records that for a signal-to-noise ratio of $r=2$ and $\eta=5\%$, we require $N_{\mathcal{O}} = 6$ instead of $N_{c}= 9$ qubits to achieve $\varepsilon < 0.08$ \cite{supplemental}. 

\emph{State and parameter estimation.$-$} Many quantum information processing applications, such as state and process tomography~\cite{haffner2005,chow2011,merkel2013,medford2013,ryan2013} and parameter estimation~\cite{shulman2014}, benefit from accurate and precise estimation of qubit observables (e.g., the Pauli operators). Analog data processing has been used extensively, e.g., for parameter \cite{wiseman1997} and state~\cite{banaszek1999} estimation in quantum optical systems, where it is often natural to process quasicontinuous field quadratures or photon counts.  For many qubit systems, the common approach is instead to threshold the data.  Thresholding the data is generally suboptimal, as we now illustrate.

For definiteness, we consider estimating the quantum expectation value $s_0=\left\langle\sigma_z\right\rangle$ of the single-qubit Pauli operator $\sigma_z$ (in the basis $\ket{\pm}$) from the independent readout of $N$ identically prepared copies of a qubit. As in the case of the repetition code, we compare the standard maximum-likelihood estimator (MLE)~\cite{cramer1946} applied to the analog data set $\left\{\mathcal{O}_i\right\}$ instead of the thresholded data set $\left\{c_i\right\}$ in order to benchmark the improvement. In both cases, the MLE is the value $s$ that maximizes the likelihood function $\mathcal{L}(s)=\prod_{i=1}^N P(\mathcal{O}_i/c_i|s)$ under the constraint $-1 \le s \le +1$. In practice, the MLE is obtained by maximizing the equivalent log-likelihood function $\ell(s) = N^{-1}\ln \mathcal{L}(s)$. The MLE is asymptotically unbiased, normally distributed, and minimizes the variance [i.e. saturates the Cram\'er-Rao bound, see Eq.~\eqref{eq:asymptoticMLEvariance}, below] for large $N$~\cite{cramer1946}. 

When the analog data are thresholded, the MLE is the (bias-corrected) thresholded average $s_{\textrm{TA}}=N^{-1}\sum_{i=1}^N c_i$ considered, e.g., in Ref.~\cite{ryan2013}. This estimate does not make use of the soft information contained in the distributions $P(\mathcal{O}|\pm)$ for reconstruction of $s_0$. In contrast, the soft-decoded estimate $s_{\textrm{SD}}$ obtained by applying the MLE to the analog data set makes full use of the distributions $P(\mathcal{O}|\pm)$. In Ref.~\cite{ryan2013}, the alternative \emph{soft average} $s_{\textrm{SA}}=N^{-1}\sum_{i=1}^N \mathcal{O}_i$ was also employed as an estimator for $s_0$, but this approach is also suboptimal~\cite{supplemental}.

We will measure the deviation of an estimate $s$ from the true value $s_0$ with the mean squared error (MSE) $\zeta$, given by the sum of the variance and of the squared bias of the estimator, $\zeta \equiv \statAverage{(s-\statAverage{s})^2} + (\statAverage{s}-s_0)^2$. Here, the statistical average $\statAverage{\,}$ is taken with respect to the distribution of outcomes:
\begin{align}
	P(\mathcal{O}/c|s_0)=\frac{1+s_0}{2}P(\mathcal{O}/c|+) + \frac{1-s_0}{2}P(\mathcal{O}/c|-). \label{eq:sDistribution}
\end{align}
For this distribution, $\ell(s)$ is a concave function with a unique maximum. A general expression for the asymptotic {MSE} of the thresholded average $s_{\textrm{TA}}$ can be derived~\cite{supplemental}. For the Gaussian readout, it takes the simple form reported in Ref.~\cite{ryan2013}, $\zeta_{\textrm{TA}} \sim [\left(1-2\epsilon\right)^{-2}-s_0^2]/N$. For the soft-decoded MLE estimate, the asymptotic MSE can be computed directly from the Fisher information of $P(\mathcal{O}|s_0)$:
\begin{align}
	\zeta_{\textrm{SD}} \sim -\frac{1}{N}\statAverage{\frac{\partial^2 \ln P(\mathcal{O}|s_0)}{\partial s_0^2}}^{-1}. \label{eq:asymptoticMLEvariance}
\end{align}
Here, we use the symbol ``$\sim$'' to indicate a strict asymptotic equality. From Eqs.~\eqref{eq:sDistribution} and \eqref{eq:asymptoticMLEvariance}, an explicit asymptotic form for $\zeta_{\textrm{SD}}$ can be found in terms of the distributions $P(\mathcal{O}|\pm)$ \cite{supplemental}:
\begin{align}
	\zeta_{\textrm{SD}}\sim \frac{1}{N} \cdot \frac{1-s_0^2}{1-I},\quad I=\int d\mathcal{O}\,\frac{P(\mathcal{O}|+)P(\mathcal{O}|-)}{P(\mathcal{O}|s_0)}. \label{eq:explicitLikelihoodMSE}
\end{align}
In Eq.~\eqref{eq:explicitLikelihoodMSE}, the integral $I$ contains all information about the noise introduced by the readout apparatus. The remaining contribution when $I=0$ is the quantum shot noise (projection noise), which reflects the choice of a particular measurement basis. Since $I$ has the form of an overlap integral, it is especially important to understand the tails of the (generally non-Gaussian) readout distributions $P(\mathcal{O}|\pm)$.

\begin{figure}
\centering
\includegraphics[width=\columnwidth]{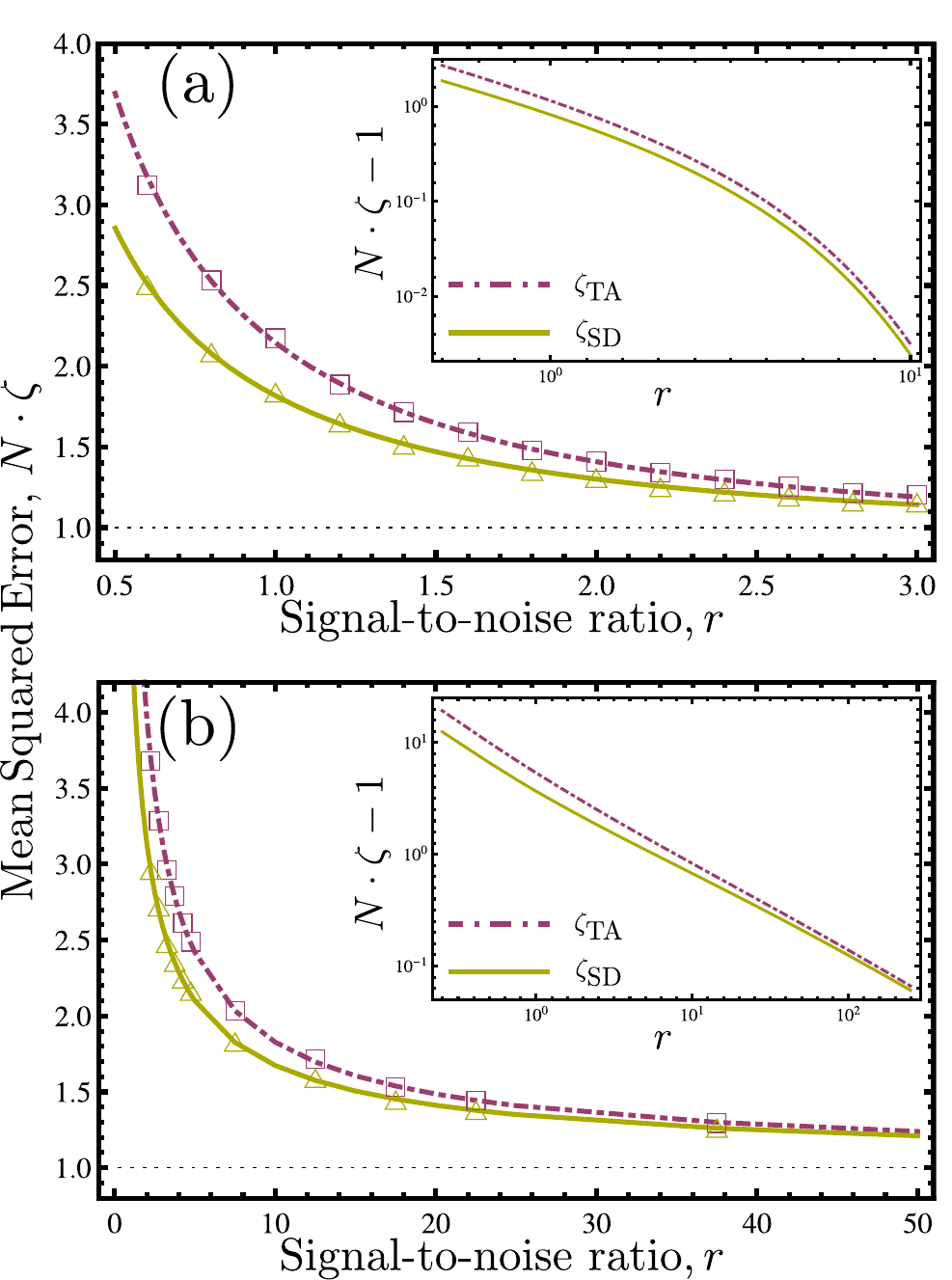}
\centering
\caption{(Color online) Asymptotic normalized MSEs $N\cdot\zeta$ of the soft-decoded estimate (solid gold), Eq.~\eqref{eq:explicitLikelihoodMSE}, and of the thresholded average (dot-dashed magenta line), given in Ref.~\cite{supplemental}, as a function of the signal-to-noise ratio $r$ assuming $s_0=\left\langle \sigma_z\right\rangle=0$ for (a) the Gaussian readout and (b) the ``peak-signal'' readout of Ref.~\cite{danjou2014}. The finite-$N$ MSEs of the thresholded average (magenta squares) and soft-decoded estimate (gold triangles) are obtained from $5\times 10^4$ randomly generated measurement records of $N=100$ qubits. Insets: Asymptotic MSEs on a logarithmic scale. \label{fig:fig2}}
\end{figure}

To quantitatively verify that soft decoding can improve state estimation, we set $s_0=0$ and evaluate Eq.~\eqref{eq:explicitLikelihoodMSE} numerically for both the Gaussian readout and the peak-signal readout of Ref.~\cite{danjou2014}. We plot the asymptotic MSE as a function of the signal-to-noise ratio $r$ in Fig.~\ref{fig:fig2}. We also plot the asymptotic MSE of the bias-corrected thresholded average, $\zeta_{\textrm{TA}}$. Figure~\ref{fig:fig2} confirms that soft decoding always outperforms thresholding (i.e., $\zeta_\mathrm{SD}<\zeta_\mathrm{TA}$). As $r$ increases, $\zeta_{\textrm{TA}}$ and $\zeta_{\textrm{SD}}$ exhibit an approximate power-law approach to the projection-noise limit for the (non-Gaussian) peak-signal readout of Ref.~\cite{danjou2014}, whereas they decrease exponentially for the Gaussian readout. In the intermediate regime for $r$, there is a clear advantage in soft decoding with the MLE for both readouts, demonstrating substantial benefits in the experimentally relevant regime of signal-to-noise ratios, $r \sim 1$ for the Gaussian readout~\cite{ryan2013}, and $r\sim 10$ for the peak-signal readout.

To show that the asymptotic advantage persists when $N$ is finite, we calculate $\zeta$ from a Monte Carlo simulation with $N=100$. We first randomly generate $5\times 10^4$ measurement records $\left\{\mathcal{O}_i\right\}$ from the distribution $\prod_{i=1}^N P(\mathcal{O}_i|s_0)$ for both the Gaussian readout and non-Gaussian peak-signal readout of Ref.~\cite{danjou2014}. For each measurement record, we calculate $s_{\textrm{TA}}$ and optimize the log-likelihood function $\ell(s)$ to obtain $s_\mathrm{SD}$. We then directly obtain $\zeta$ from the variance and bias of the simulated estimates. The results are displayed (open symbols) in Fig.~\ref{fig:fig2} for $s_0=0$. The simulated data points coincide with the asymptotic predictions.

In conclusion, we have shown that making use of the soft information contained in the analog outcomes of a qubit readout apparatus, as opposed to irreversibly thresholding each qubit to a binary value, can significantly improve the performance of quantum information processing tasks involving the measurement of many qubits. We have focused on two examples of practical importance. In the case of enhanced state detection with the quantum repetition code, the number of qubits required to achieve a given error rate can be reduced by up to a factor of $2$ through improved data processing alone. Importantly, we have shown that an advantage persists for small numbers of qubits and finite encoding errors. In addition, we have shown that optimal processing of analog qubit readout outcomes can appreciably increase the precision on the measurement of qubit observables (e.g., the Pauli operators). Crucially, in both cases we have demonstrated a significant improvement for an experimentally relevant, non-Gaussian qubit readout model~\cite{elzerman2004,morello2010,pla2013,veldhorst2014,danjou2014}.

Our results offer encouraging prospects for direct improvements of both small and large scale quantum information processing applications through soft decoding of the qubit readout. For example, readout error models for decoders of topological codes \cite{duclos-cianci2010,fowler2012} could be modified using the ideas presented here to accept analog readout outcomes with realistic statistics at the single-physical-qubit level, improving error detection rates. While there are many possible extensions of this work, the direct improvements we have shown to enhanced state detection and quantum state or parameter estimation are both practical and immediately realizable in a wide array of current experiments.

\nocite{rubinstein2008,press1992}

We thank L. Childress, A. Fowler, and D. Poulin for useful discussions. We acknowledge financial support from the National Sciences and Engineering Research Council of Canada (NSERC), the Canadian Institute for Advanced Research (CIFAR), the Fonds the Recherche du Qu\'ebec Nature et Technologies (FRQNT), and the Institut Transdisciplinaire d'Information Quantique (INTRIQ).

%


\clearpage

\renewcommand{\theequation}{S.\arabic{equation}}
\setcounter{equation}{0}

\onecolumngrid

\section*{\Large Supplemental Material}

\section{Realistic non-Gaussian readout: the peak-signal readout}

To show that soft decoding of a readout leads to an advantage for realistic readouts, we have investigated the performance of soft decoding for the experimentally relevant~\cite{elzerman2004,morello2010,veldhorst2014} `peak-signal' readout analyzed in Ref.~\cite{danjou2014}. In this section, we briefly summarize this readout for completeness.
\begin{suppfigure}
\centering
\includegraphics[width=\columnwidth]{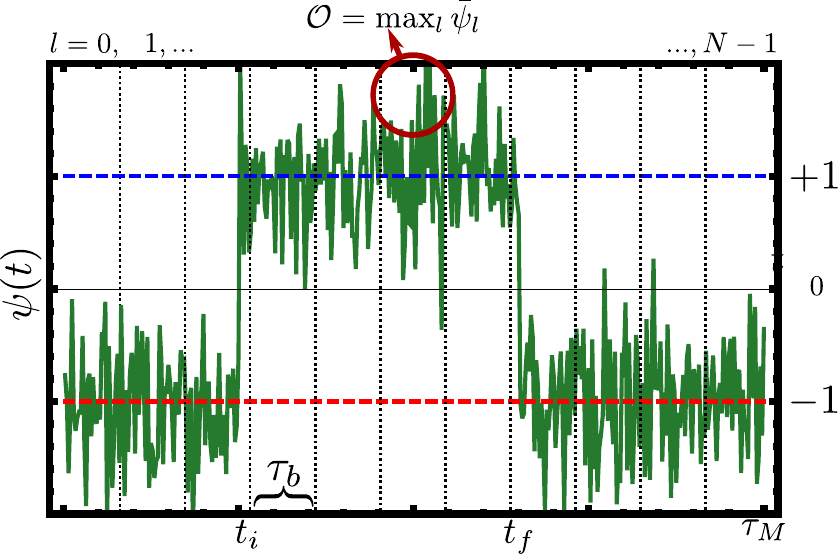}
\centering
\caption{(Color online) Generic time-dependent signal $\psi(t)$ signalling the excited state $\ket{+}$ for the class of readouts discussed in Ref.~\cite{danjou2014}. The turn-on time $t_i$ and pulse width $t_f-t_i$ both follow Poisson statistics. The measurement time $\tau_M$ is binned into subintervals of length $\tau_b$ and the observable $\mathcal{O}$ is chosen to be the maximum of the signal over all bins. The parameters $\tau_M$ and $\tau_b$ are chosen to minimize the single-shot error rate. \label{fig:figS1}}
\end{suppfigure}

In this readout, the ground and excited states $\ket{-}$ and $\ket{+}$ are mapped to a time-dependent signal $\psi(t)$ subject to Gaussian white noise. When the state is $\ket{-}$, the signal has a constant value $-1$ on average. When the state is $\ket{+}$, however, the signal is a square pulse starting at a random turn-on time $t_i$ and ending at a random turn-off time $t_f$, as illustrated in Fig.~S.\ref{fig:figS1}. The times $t_i$ and $t_f-t_i$ each follow their own Poisson statistics. The measurement time $\tau_M$ is divided into $N$ bins of length $\tau_b$, with the average signal on the $l^{\textrm{th}}$ bin being $\bar{\psi}_l$. The observable $\mathcal{O}$ is then chosen to be the maximum of $\bar{\psi}_l$ over all bins. Finally, the measurement time $\tau_M$, bin time $\tau_b$ and threshold $\nu$ are chosen to optimize the single-shot readout fidelity. Other choices for the observable $\mathcal{O}$ have also been discussed in Refs.~\cite{gambetta2007,danjou2014}.

Typical probability distributions $P(\mathcal{O}|\pm)$ for the peak-signal readout are shown in Fig.~S.\ref{fig:figS2} for two different values of the power signal-to-noise ratio $r$ (integrated over a time $\left\langle t_f - t_i \right\rangle$). We see that the distributions have prominent non-Gaussian features. To perform fast sampling of these distributions, we first cut off the tails of each distribution (the lost probability weight is smaller than about $10^{-7}$) and renormalize them. Next, we numerically integrate the analytic expressions for $P(\mathcal{O}|\pm)$ given in Ref.~\cite{danjou2014} to construct a linear interpolation of the inverse cumulative distribution function $Q_{\pm}^{-1}$ associated with $P(\mathcal{O}|\pm)$, as shown in Fig.~S.\ref{fig:figS2}. An independent sample of $P(\mathcal{O}|\pm)$ is then given by $\mathcal{O}=Q_{\pm}^{-1}(x)$, where $x$ is a random number generated from a uniform distribution between $0$ and $1$~\cite{rubinstein2008}.

\begin{suppfigure}
\centering
\includegraphics[width=\columnwidth]{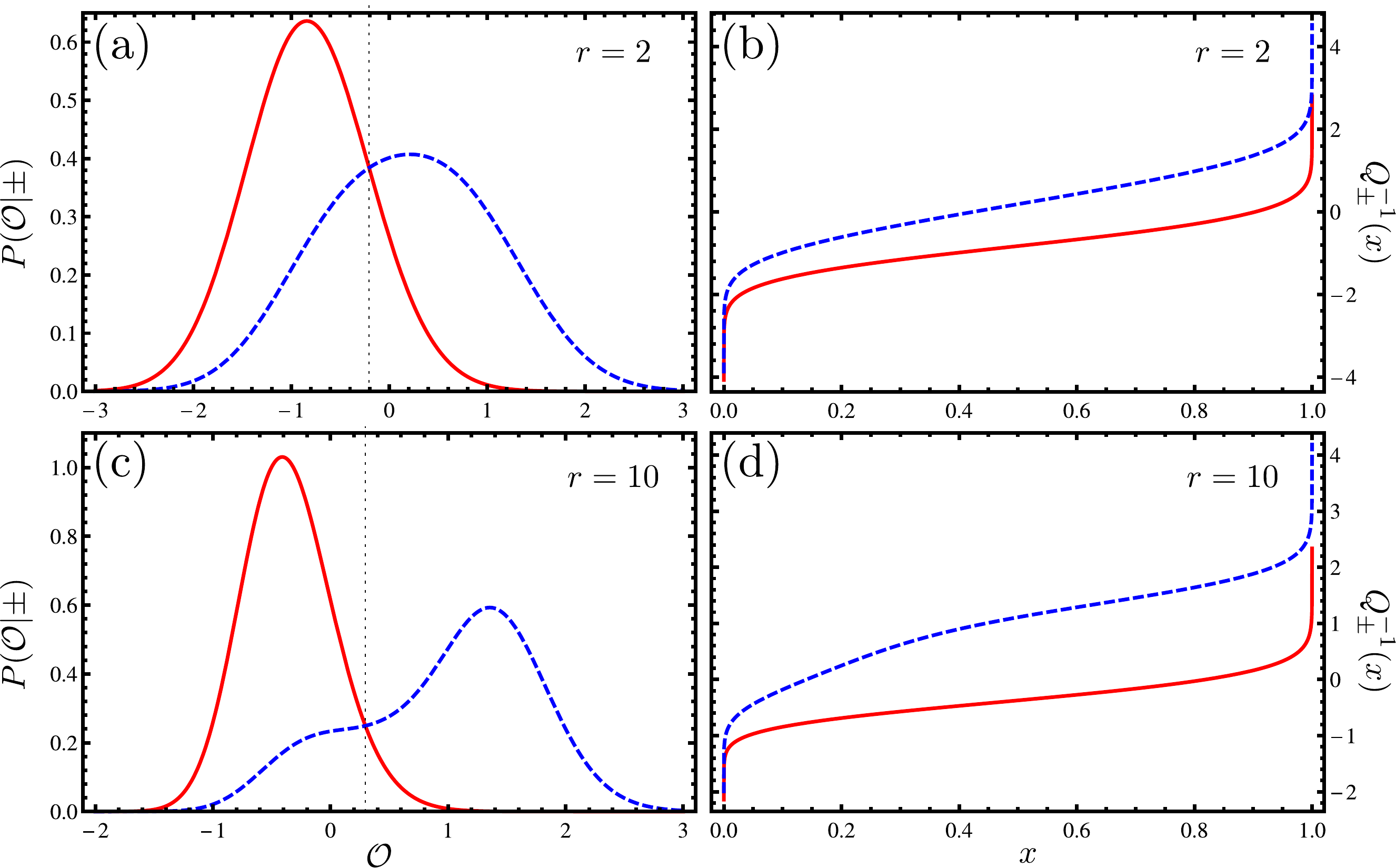}
\centering
\caption{(Color online) Example of optimal peak-signal distributions $P(\mathcal{O}|\pm)$ given in Ref.~\cite{danjou2014} for a signal-to-noise ratio of (a) $r=2$ and (c) $r=10$. The dotted vertical line is the optimal threshold. In both cases, $\langle t_f-t_i\rangle/\langle t_i\rangle = 4$. The corresponding inverse cumulative distribution functions $Q_{\pm}^{-1}(x)$ used for fast sampling of the distributions are shown in panels (b) and (d). \label{fig:figS2} }
\end{suppfigure}

\section{Enhanced state detection with the quantum repetition code}

\subsection{Advantage of soft decoding}

In this section, we give a brief derivation of the advantage obtained for enhanced state detection through soft decoding of the quantum repetition code for a Gaussian readout, Eq.~(4) of the main text. Note that to benefit from enhanced state detection, coherent encoding of the logical state $\ket{\psi} = \alpha_0 \ket{0} + \alpha_1 \ket{1}$ into the state $\ket{\psi_N} = \alpha_0 \ket{-}^{\otimes N} + \alpha_1 \ket{+}^{\otimes N}$ is not necessary. For example, encoding into the mixed state $\rho_N = |\alpha_0|^2 \ketbra{-}{-}^{\otimes N} + |\alpha_1|^2 \ketbra{+}{+}^{\otimes N}$ (e.g. by allowing the qubits to purely dephase) gives the same advantage.

For the Gaussian readout, each qubit measurement yields an analog value $\mathcal{O}$ conditioned on the qubit state $\ket{\pm}$ according to the probability distributions:
\begin{align}
	P(\mathcal{O}|\pm) = \sqrt{\frac{r}{2\pi}}e^{-\frac{(\mathcal{O}\mp 1)^2 r}{2}}. \label{eq:gaussianDistributions}
\end{align}
Here, $r$ is the power signal-to-noise ratio and the signal is normalized to have average values $\pm 1$. If the qubit is read out in a single shot, each analog outcome $\mathcal{O}_i$ is converted to a binary outcome $c_i = c_{\pm}$ by setting a threshold $\nu$. The single-shot error rates conditional on the qubit state are:
\begin{align}
	\epsilon_- = P(c_+|-) = \int_{\nu}^\infty d\mathcal{O}\,P(\mathcal{O}|-),\;\;\epsilon_+ = P(c_-|+) = \int_{-\infty}^\nu d\mathcal{O}\,P(\mathcal{O}|+).
\end{align} 
Assuming equal \emph{a priori} probabilities for the ground and excited states, the average single-shot readout error rate $\epsilon=(\epsilon_+ + \epsilon_-)/2$ is minimized by choosing $P(\nu|+)=P(\nu|-)\Rightarrow \nu=0$. An explicit calculation of the integrals gives:
\begin{align}
	\epsilon = \epsilon_{\pm} = \frac{1}{2} \mathrm{erfc}\left(\sqrt{\frac{r}{2}}\right). \label{eq:singleShotErrorRate}
\end{align}
Eq.~\eqref{eq:singleShotErrorRate} defines the binary symmetric readout associated with the Gaussian readout.

We assume for simplicity that when the $N$ qubits of the quantum repetition code are measured, the $N$-qubits state collapses to $\ket{+}^{\otimes N}$ or $\ket{-}^{\otimes N}$ with equal probability. The resulting dataset consists of $N$ analog readout outcomes $\mathcal{O}_i$. In the main text, we gave two likelihood ratios $\Lambda_{c}$ and $\Lambda_{\mathcal{O}}$ for thresholded and analog readout outcomes, respectively. In both cases, if $\Lambda > 1$ ($\Lambda < 1$) we infer that the qubit state is $\ket{1}$ ($\ket{0}$). For the Gaussian readout, the likelihood ratios reduce to:
\begin{align}
	\Lambda_{c} = \left(\frac{1-\epsilon}{\epsilon}\right)^{2n_+ - N}, \;\;\;\; \Lambda_{\mathcal{O}} = \exp\left(2 N r \bar{\mathcal{O}}\right), \label{eq:estimatorsBSC}
\end{align} 
where $n_+$ is the number of times that the outcome $\mathcal{O}_i$ is converted to $c_+$ if the qubits are read out in a single shot and where $\bar{\mathcal{O}}=N^{-1}\sum_{i=1}^N \mathcal{O}_i$ is the sample average of the analog outcomes~\footnote{In Ref.~\cite{schaetz2005}, a two-qubit repetition code was implemented using two trapped ions. The enhanced detection was obtained by collecting the total fluorescence $\bar{\mathcal{O}}$ of both ions. We note that according to Eq.~\eqref{eq:estimatorsBSC}, this effectively implements soft decoding of the repetition code if the fluorescence counts follow a Gaussian readout distribution. For a general readout, however, knowledge of $\bar{\mathcal{O}}$ is not sufficient to determine $\Lambda_\mathcal{O}$; in this case the analog observable $\mathcal{O}_i$ must be recorded for each qubit.}.

Since $\epsilon < 1/2$, the likelihood ratio $\Lambda_c$ is equivalent to majority vote decoding of the repetition code. The corresponding average error rate $\varepsilon_c$ is given by the probability that $n_+ > N/2$ given that $\ket{0}$ is encoded, which is the same as the probability that $n_+ < N/2$ given that $\ket{1}$ is encoded. If $N=2M-1$ is odd, $\varepsilon_c$ is given by:
\begin{align}
	\varepsilon_c = \sum_{n_+ = \frac{N+1}{2}}^{N}\binom{N}{n_+} \epsilon^{n_+} (1-\epsilon)^{N-n_+}=I_{\epsilon}\left(\frac{N+1}{2},\frac{N+1}{2}\right). \label{eq:majorityVoteErrorRate}
\end{align}
Here, $I_{\epsilon}(a,b)$ is the regularized incomplete beta function~\cite{press1992}. The error rate for $N=2M$ is the same since the case $n_+=N/2$ provides no information on the qubit state for a binary symmetric readout. For $r$ large enough ($N \epsilon \ll 1$), Eq.~\eqref{eq:majorityVoteErrorRate} takes the approximate form:
\begin{align}
	\varepsilon_c \simeq \binom{N}{\frac{N+1}{2}}\frac{1}{\left(2\pi r\right)^{\frac{N+1}{4}}} e^{-\frac{(N+1)r}{4}}. \label{eq:asymptoticThresholdErrorRate}
\end{align}
Eq.~\eqref{eq:asymptoticThresholdErrorRate} must be contrasted to the error rate for the likelihood ratio $\Lambda_{\mathcal{O}}$ in Eq.~\eqref{eq:estimatorsBSC}. The corresponding average error rate $\varepsilon_\mathcal{O}$ is given by the probability that $\bar{\mathcal{O}}>0$ given that $\ket{0}$ is encoded, which is the same as the probability that $\bar{\mathcal{O}} < 0$ given that $\ket{1}$ is encoded. Since $P(\bar{\mathcal{O}}|1)$ and $P(\bar{\mathcal{O}}|0)$ are also Gaussians centered at $\pm 1$ with signal-to-noise ratio $N r$, the average error rate for the soft decoding of the readout apparatus is simply:
\begin{align}
	\varepsilon_{\mathcal{O}} = \int_{0}^{\infty} d\bar{\mathcal{O}}\,P(\bar{\mathcal{O}}|0)=\frac{1}{2}\mathrm{erfc}\left(\sqrt{\frac{N r}{2}}\right). \label{eq:analogErrorRate}
\end{align}
When $r \gg 1$, Eq.~\eqref{eq:analogErrorRate} becomes:
\begin{align}
	\varepsilon_\mathcal{O} \simeq \frac{1}{\sqrt{2 \pi N r}}e^{-\frac{N r}{2}}. \label{eq:asymptoticAnalogErrorRate}
\end{align}
Inspection of Eqs.~\eqref{eq:asymptoticThresholdErrorRate} and \eqref{eq:asymptoticAnalogErrorRate} suggests that $\varepsilon_{\mathcal{O}}$ decreases at approximately twice the rate of $\varepsilon_c$ when $N$ increases. Indeed, let $N_c$ and $N_\mathcal{O}$ be the number of qubits required to achieve a target error rate $\varepsilon$, i.e. $\varepsilon_c(r,N_c)=\varepsilon_{\mathcal{O}}(r,N_{\mathcal{O}})$. Using Eqs.~\eqref{eq:majorityVoteErrorRate} and \eqref{eq:analogErrorRate}, we solve this equation for $N_\mathcal{O}$ to subleading order in $r \gg N_c$ and obtain the result discussed in the main text:
\begin{align}
	N_{\mathcal{O}} = \frac{N_c+1}{2} + \frac{N_c - 1}{2}\frac{\ln r}{r} + O\left(\frac{N_c}{r}\right).
\end{align}
This expression is valid for any odd $N_c \ge 1$. The asymptotic advantage, $N_\mathcal{O} \sim N_c/2$, has been discussed for arbitrary block codes in Ref.~\cite{Chase1972}.

\subsection{Encoding errors}

In this section, we expand on the effect of encoding errors on the repetition code. As discussed in the main text, we consider only uncorrelated bit flip errors for simplicity. We consider both the Gaussian readout and the realistic peak-signal readout analyzed in Ref.~\cite{danjou2014} and summarized above.

Let $\eta$ be the probability for any qubit of the code to flip during the encoding sequence. The likelihood ratio for analog readout outcomes takes the modified form:
\begin{align}
	\Lambda_\mathcal{O} \equiv \prod_{i=1}^N \Lambda_{\mathcal{O},i}, \label{eq:analogLikelihoodRatioBitFlips}
\end{align}
where $\Lambda_{\mathcal{O},i}$ is the likelihood ratio for a single qubit measurement:
\begin{align}
	\Lambda_{\mathcal{O},i}= \frac{P(\mathcal{O}_i|1)}{P(\mathcal{O}_i|0) } = \frac{(1-\eta) P(\mathcal{O}_i|+) + \eta P(\mathcal{O}_i|-)}{(1-\eta)P(\mathcal{O}_i|-) + \eta P(\mathcal{O}_i|+) }. \label{eq:singleMeasurementLikelihoodRatio}
\end{align}
If the \emph{a priori} probabilities of the logical states $\ket{0}$ and $\ket{1}$ are equal, the single-shot threshold $\nu$ is obtained as usual from $P(\nu|1)=P(\nu|0)\,\Rightarrow\,P(\nu|+)=P(\nu|-)$. Therefore, sufficiently localized readout probability distributions such as the Gaussian distributions satisfy:
\begin{align}
	\Lambda_{\mathcal{O},i} \approx \frac{1-\eta}{\eta} \;\;\;\; \left(\mathcal{O} \gg \nu\right), \;\;\;\;\;\; \Lambda_{\mathcal{O},i} \approx \frac{\eta}{1-\eta} \;\;\;\; \left(\mathcal{O} \ll \nu\right).
\end{align}
Because $\Lambda_{\mathcal{O},i}$ is approximately constant above and below threshold, it seems that soft decoding of the analog readout outcomes is reduced to a simple thresholding procedure when $\eta$ is finite. As illustrated in Fig.~S.\ref{fig:figS3}, additional information can nevertheless be extracted from values $\mathcal{O}_i$ falling near the threshold, where $\Lambda_{\mathcal{O},i}$ is non-constant, provided that $\eta$ is small enough.
\begin{suppfigure}
\centering
\includegraphics[width=\columnwidth]{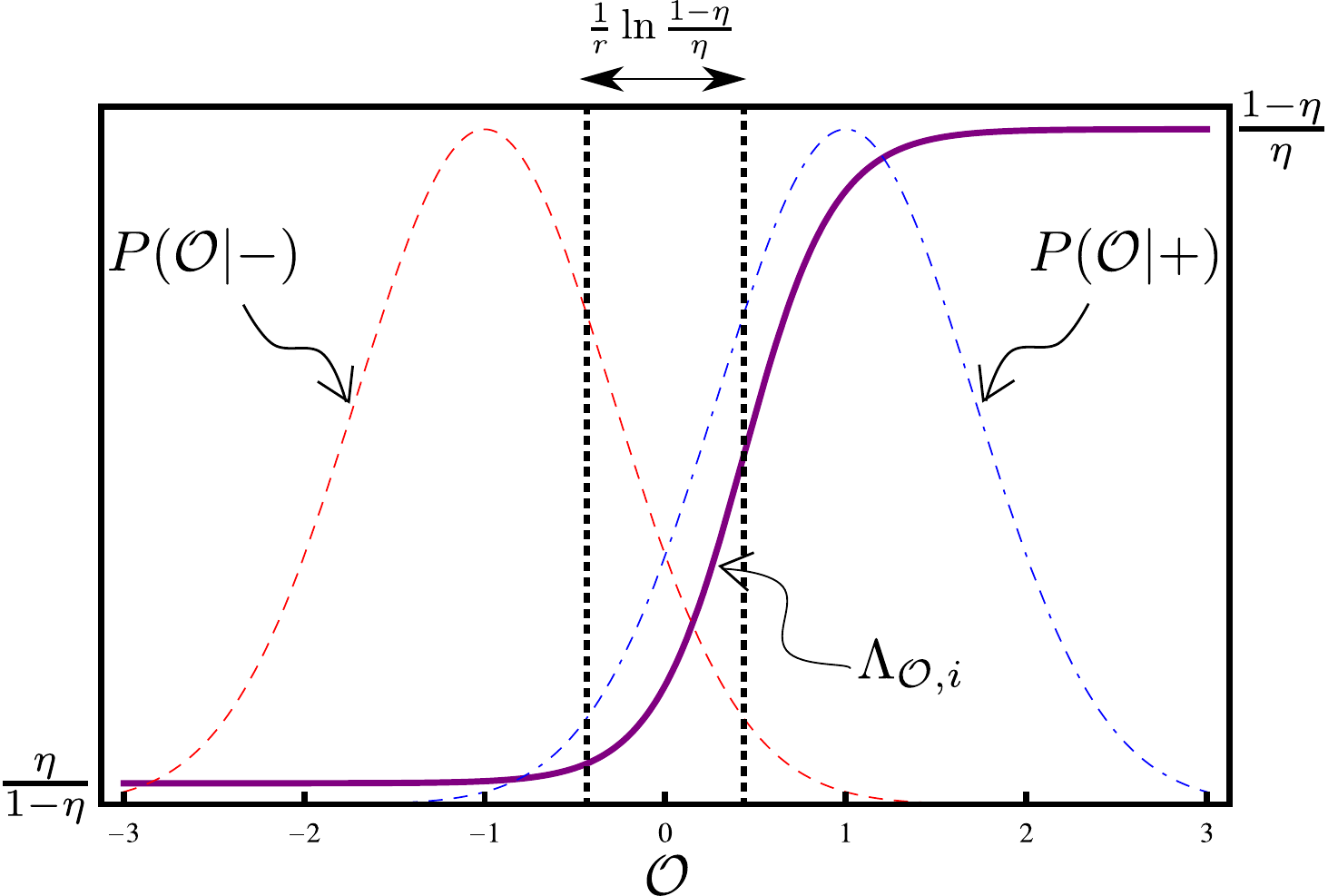}
\centering
\caption{(Color online) Schematic plot of the likelihood ratio $\Lambda_{\mathcal{O},i}$ for a single qubit measurement, Eq.~\eqref{eq:singleMeasurementLikelihoodRatio}, as a function of $\mathcal{O}$ for the Gaussian readout with $r=2$ (the values on the vertical axis have been rescaled for clarity). Far above (below) threshold, $\Lambda_{\mathcal{O},i}$ is approximately constant. However, $\Lambda_{\mathcal{O},i}$ is non-constant on an interval of width $\frac{1}{r}\ln \frac{1-\eta}{\eta}$. \label{fig:figS3}}
\end{suppfigure}

In the case of the Gaussian readout distributions, $\Lambda_{\mathcal{O},i}$ is non-constant for values of $\mathcal{O}$ such that $\exp\left(-2 r|\mathcal{O}|\right) \gtrsim \eta/(1-\eta)$, as illustrated in Fig.~S.\ref{fig:figS3}. In order for a significant fraction of measured values to lie in that interval, we must have $|\mathcal{O}| \gtrsim 1$. Therefore:
\begin{align}
	\frac{\eta}{1-\eta} \lesssim e^{-2 r}\;\;\; \Rightarrow \;\;\; \eta \lesssim \frac{e^{-2 r}}{1+e^{-2 r}}.
\end{align}
In the limit $r \gg 1$, this reduces to $\eta \lesssim e^{-2r}$. Since Eq.~\eqref{eq:singleShotErrorRate} implies that $\epsilon \sim e^{-\frac{r}{2}}$ up to logarithmic corrections for $r \gg 1$, we conclude that $\eta$ must be smaller than some power of $\epsilon$. As shown in Fig.~S.\ref{fig:figS5}, a similar upper bound on $\eta$ exists for the non-Gaussian peak-signal readout of Ref.~\cite{danjou2014}.

If each analog outcome $\mathcal{O}_i$ is instead thresholded to a binary outcome $c_i=c_\pm$, the likelihood ratio is:
\begin{align}
	\Lambda_c = \left[\frac{(1-\eta)(1-\epsilon_+)+\eta \epsilon_-}{(1-\eta)\epsilon_- + \eta (1-\epsilon_+)}\right]^{n_+}\cdot \left[\frac{(1-\eta)\epsilon_+ +\eta (1-\epsilon_-)}{(1-\eta)(1-\epsilon_-) + \eta \epsilon_+}\right]^{N-n_+}, \label{eq:singleShotLikelihoodRatioBitFlips}
\end{align}
where $n_+$ is the number of qubits that are assigned the value $c_+$. To show quantitatively that a significant advantage can be obtained by utilizing the analog readout outcomes in the presence of encoding errors, we must compare the performance of Eq.~\eqref{eq:singleShotLikelihoodRatioBitFlips} to that of Eq.~\eqref{eq:analogLikelihoodRatioBitFlips}.

For both the Gaussian readout and the realisitic peak-signal readout, we performed Monte-Carlo simulations of the error rates for maximum-likelihood decoding of the analog and thresholded readout outcomes. In both cases, we take the signal-to-noise ratio to be $r=2$ and choose parameters that optimize the single-shot readout fidelity. We randomly choose the logical state $\ket{0}$ or $\ket{1}$ with equal probability and generate a random measurement record by sampling $N$ independent values from the distributions $P(\mathcal{O}_i|1)=(1-\eta) P(\mathcal{O}_i|+) + \eta P(\mathcal{O}_i|-)$ or $P(\mathcal{O}_i|0)=(1-\eta) P(\mathcal{O}_i|-) + \eta P(\mathcal{O}_i|+)$, respectively. We then infer the state with both Eq.~\eqref{eq:analogLikelihoodRatioBitFlips} and Eq.~\eqref{eq:singleShotLikelihoodRatioBitFlips} and record an error if the decision is incorrect. We repeat the procedure $10^7$ times ($10^6$ times) for the Gaussian readout (peak-signal readout) and obtain the error rate from the ratio of errors to the number of trials. The resulting error rates are shown in Figs.~S.\ref{fig:figS4} and S.\ref{fig:figS5} respectively. For the Gaussian readout with $\eta = 0$, we instead plot the analytic expressions, Eqs.~\eqref{eq:majorityVoteErrorRate} and \eqref{eq:analogErrorRate}. Numerical values of the error rates for both readouts without encoding errors are tabulated in Table S.\ref{tab:tabS1} for convenience.

\begin{suppfigure}
\centering
\includegraphics[width=\columnwidth]{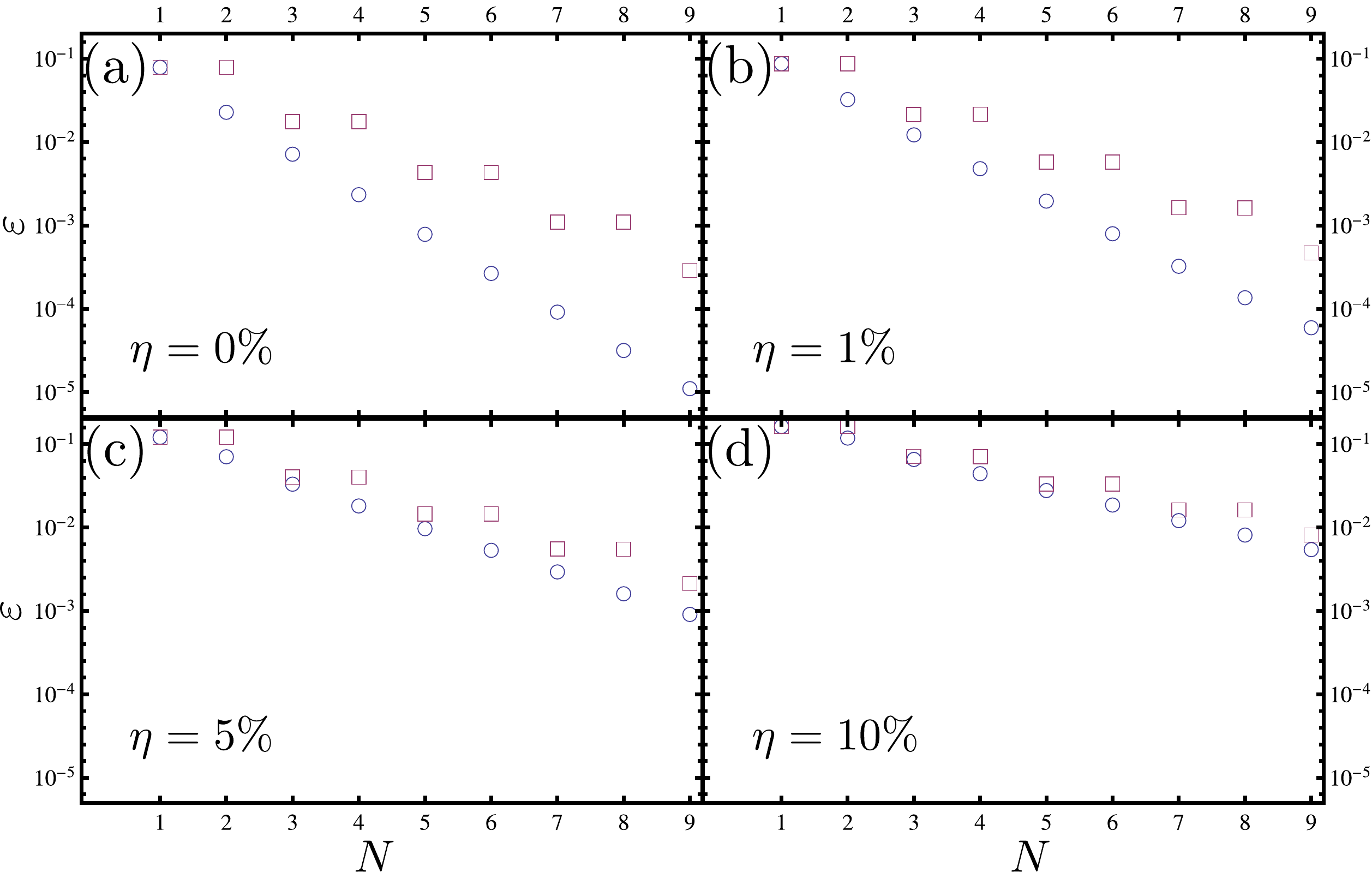}
\centering
\caption{(Color online) Simulated repetition code error rates $\varepsilon$ as a function of the number of qubits $N$ for the Gaussian readout with signal-to-noise ratio $r=2$. The error rate was obtained for soft decoding (blue circle), Eq.~\eqref{eq:analogLikelihoodRatioBitFlips}, and thresholding (magenta square), Eq.~\eqref{eq:singleShotLikelihoodRatioBitFlips}, of the readout outcomes. The error rate for thresholding is the same for $N=2M$ as for $N=2M-1$ since the case $n_+ = N/2$ provides no information on the qubit state ($\epsilon_+ = \epsilon_- = \epsilon$). Each panel corresponds to a different encoding error rate $\eta$. For $\eta=0$, we plotted Eqs.~\eqref{eq:majorityVoteErrorRate} and \eqref{eq:analogErrorRate}. For $\eta\neq 0$, the error rates were calculated by generating $10^7$ random measurement records $\left\{\mathcal{O}_i\right\}$ sampled with equal probability from $P(\left\{\mathcal{O}_i\right\}|1)$ and $P(\left\{\mathcal{O}_i\right\}|0)$.\label{fig:figS4}}
\end{suppfigure}

\begin{suppfigure}
\centering
\includegraphics[width=\columnwidth]{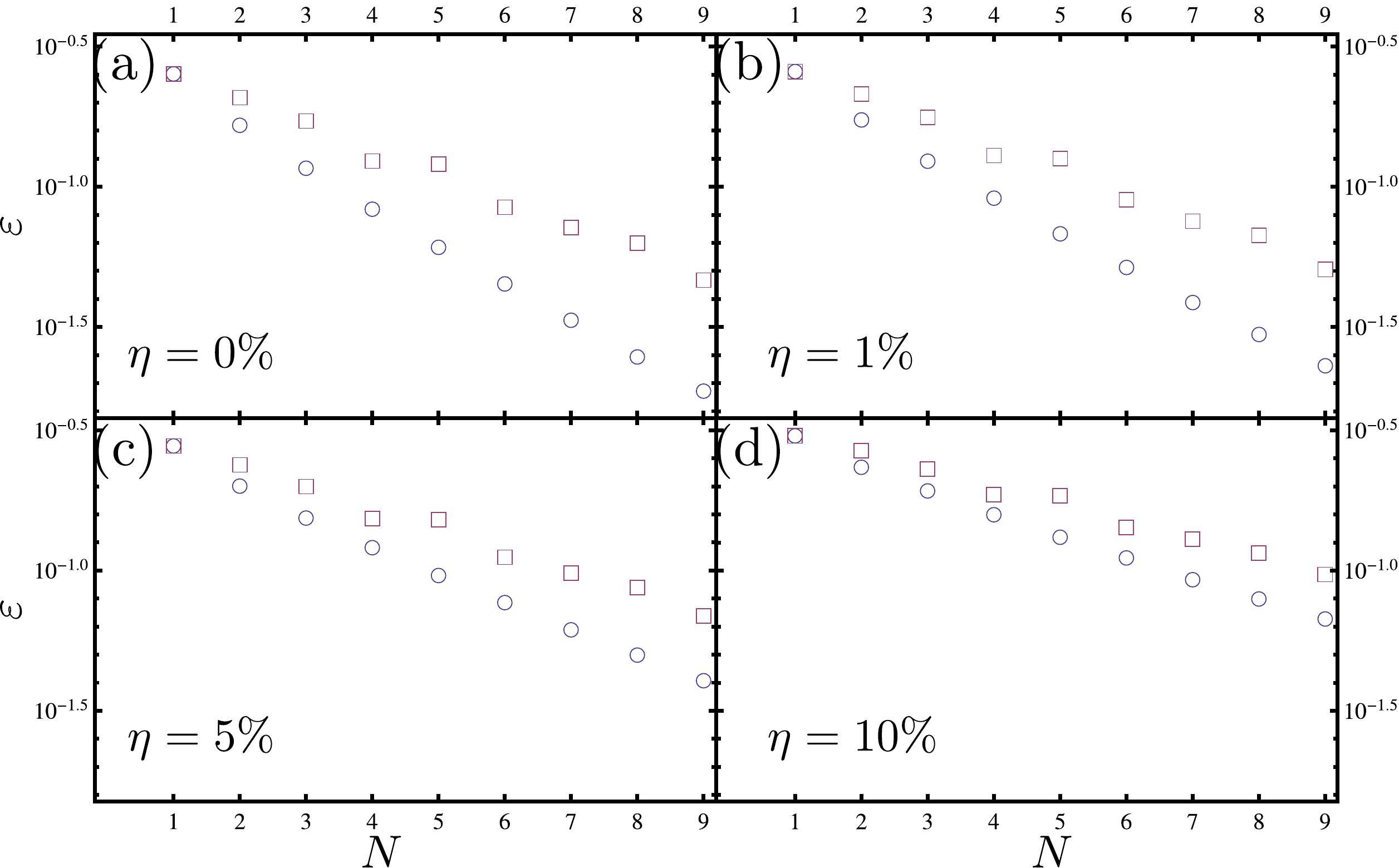}
\centering
\caption{(Color online) Simulated repetition code error rates $\varepsilon$ as a function of the number of qubits $N$ for the peak-signal readout of Fig.~S.\ref{fig:figS1} with signal-to-noise ratio $r=2$ and $\langle t_f-t_i\rangle/\langle t_i\rangle = 4$. The error rate was obtained for soft decoding (blue circle), Eq.~\eqref{eq:analogLikelihoodRatioBitFlips}, and thresholding (magenta square), Eq.~\eqref{eq:singleShotLikelihoodRatioBitFlips}, of the readout outcomes. The error rate for thresholding follows a jagged pattern since the case $n_+ = N/2$ only gives partial information on the qubit state ($\epsilon_+\neq \epsilon_-$). Each panel corresponds to a different encoding error rate $\eta$. The error rates were calculated by generating $10^6$ random measurement records $\left\{\mathcal{O}_i\right\}$ sampled with equal probability from $P(\left\{\mathcal{O}_i\right\}|1)$ and $P(\left\{\mathcal{O}_i\right\}|0)$.\label{fig:figS5}}
\end{suppfigure}

\begin{center}
\begin{supptable}
\begin{tabular}{| c || c | c | c | c | c | c | c | c | c |}
\hline
\multicolumn{10}{|c|}{{\bf Gaussian readout}} \\ \hline
$N$ & 1 & 2 & 3 & 4 & 5 & 6 & 7 & 8 & 9 \\ \hline \hline
$\varepsilon_c \, (10^{-2})$ & 7.86 & 7.86 & 1.76 & 1.76 & 0.431 & 0.431 & 0.110 & 0.110 & 0.0289 \\ \hline
$\varepsilon_{\mathcal{O}} \, (10^{-2})$ & 7.86 & 2.28 & 0.715 & 0.234 & 0.0783 & 0.0266 & 0.00914 & 0.00317 & 0.00110  \\ \hline
\multicolumn{10}{|c|}{{\bf Peak-signal readout}} \\ \hline
$N$ & 1 & 2 & 3 & 4 & 5 & 6 & 7 & 8 & 9 \\ \hline \hline
$\varepsilon_c$ & 0.253 & 0.208 & 0.172 & 0.124 & 0.121 & 0.0845 & 0.0715 & 0.0630 & 0.0465 \\ \hline
$\varepsilon_{\mathcal{O}}$ & 0.253 & 0.166 & 0.116 & 0.0832 & 0.0608 & 0.0450 & 0.0334 & 0.0247 & 0.0187 \\ \hline
\end{tabular}
\caption{Tabulated values of the thresholded and soft-decoded error rates $\varepsilon_{c}$ and $\varepsilon_\mathcal{O}$ for different numbers $N$ of repetition code qubits for both the Gaussian and peak-signal readouts. In both cases, the signal-to-noise ratio is $r=2$ and there are no encoding errors, $\eta=0$. These values correspond to those plotted in Figs.~S.\ref{fig:figS4} and S.\ref{fig:figS5}. \label{tab:tabS1}}
\end{supptable}
\end{center}

\section{State and parameter estimation}

In this section, we give a brief derivation of the asymptotic mean squared error of the maximum-likelihood estimator for $s_0 = \left\langle \sigma_z \right\rangle$ when applied to analog and thresholded readout outcomes. We also review the soft average discussed in Ref.~\cite{ryan2013}. In all cases, we estimate $s_0$ with $N$ independent analog or thresholded readout outcomes, $\left\{\mathcal{O}_i\right\}$ or $\left\{c_i\right\}$, following a distribution of the form:
\begin{align}
	P(\mathcal{O}_i/c_i|s_0) = \frac{1+s_0}{2} P(\mathcal{O}_i/c_i|+) + \frac{1-s_0}{2} P(\mathcal{O}_i/c_i|-). \label{eq:distributionS0}
\end{align}
In the following, we will denote statistical expectation values with respect to Eq.~\eqref{eq:distributionS0} by the double brackets $\statAverage{\,}$. The maximum-likelihood estimator is the value $s$ that maximizes the log-likelihood function:
\begin{align}
	\ell(s) = \frac{1}{N}\sum_{i=1}^N \ln P(\mathcal{O}_i/c_i|s), \label{eq:likelihoodFunction}
\end{align}
under the constraint $-1 \le s \le 1$.

\subsection{Thresholded readout outcomes}

First we assume that the values $\mathcal{O}_i$ are thresholded to a binary outcome $c_\pm$, where the threshold $\nu$ is chosen to satisfy $P(\nu|+)=P(\nu|-)$.

To obtain the maximum-likelihood estimator, we must maximize the likelihood function, Eq.~\eqref{eq:likelihoodFunction}. We first note that Bayes' rule gives the probability of an outcome $c_i$ given the true expectation $s_0$:
\begin{align}
	P(c_i|s_0) = \frac{1+s_0}{2} P(c_i|+) + \frac{1-s_0}{2} P(c_i|-). \label{eq:outcomeGivenS0}
\end{align}
Here, the transition probabilities of the binary readout are given by the conditional single-shot error rates:
\begin{align}
	P(c_-|+) \equiv \epsilon_+ = \int_{-\infty}^{\nu} d\mathcal{O}\, P(\mathcal{O}|+),\;\; P(c_+|-) \equiv \epsilon_- = \int_{\nu}^{\infty} d\mathcal{O}\, P(\mathcal{O}|-).
\end{align}
Thus, Eq.~\eqref{eq:outcomeGivenS0} becomes:
\begin{align}
	P(c_+|s_0) = \frac{1+s_0}{2} (1-\epsilon_+) + \frac{1-s_0}{2} \epsilon_-,\;\;P(c_-|s_0) = \frac{1+s_0}{2} \epsilon_+ + \frac{1-s_0}{2} (1-\epsilon_-). \label{eq:outcomeGivenS0Explicit}
\end{align}
Next, we use the form of Eq.~\eqref{eq:outcomeGivenS0Explicit} in the log-likelihood function, Eq.~\eqref{eq:likelihoodFunction}, and optimize with respect to $s$. Maximizing without the constraint $-1 \le s \le 1$ (i.e. setting $d\ell(s)/ds=0$), the optimum is the thresholded average:
\begin{align}
	s_{\textrm{TA}} = \frac{1}{N}\sum_{i=1}^{N} c_i, \label{eq:thresholdedAverage}
\end{align}
where the binary outcomes $c_i = c_{\pm}$ are chosen to be:
\begin{align}
	c_+ = \frac{1+(\epsilon_+ - \epsilon_-)}{1-(\epsilon_+ + \epsilon_-)},\;\;c_- = - \frac{1-(\epsilon_+ - \epsilon_-)}{1-(\epsilon_+ + \epsilon_-)}. \label{eq:unbiasedOutcomes}
\end{align}
In the limit of large $N$, the estimate is unlikely to fall outside the region $-1 \le s \le 1$. In this asymptotic limit, the estimate is unbiased:
\begin{align}
	\statAverage{s_{\textrm{TA}}} &= P(c_+|s_0) c_+ + P(c_-|s_0) c_- \\
	&= \frac{1+s_0}{2}\left[(1-\epsilon_+) c_+ + \epsilon_+ c_-\right]+\frac{1-s_0}{2}\left[\epsilon_- c_+ + (1-\epsilon_-) c_-\right] = s_0.
\end{align}
In this case, the asymptotic mean squared error $\zeta_{\textrm{TA}}$ of the maximum-likelihood estimate, Eq.~\eqref{eq:thresholdedAverage}, is equal to its asymptotic variance and is given by the central limit theorem:
\begin{align}
	\zeta_{\textrm{TA}} = \statAverage{\Delta s_{\textrm{TA}}^2} \sim \frac{\statAverage{ \Delta c^2 }}{N}=\frac{P(c_+|s_0)c_+^2 + P(c_-|s_0)c_-^2 - s_0^2}{N}. \label{eq:thresholdedMSE}
\end{align}
In the special case of a binary symmetric readout with $\epsilon_+ = \epsilon_-= \epsilon$, we have $c_+ = -c_- = (1-2\epsilon)^{-1}$ and we recover the expression given in Ref.~\cite{ryan2013}:
\begin{align}
	\zeta_{\textrm{TA}} \sim \frac{(1-2\epsilon)^{-2}-s_0^2}{N}.
\end{align}

\subsection{Analog readout outcomes}

The asymptotic mean squared error $\zeta_{\textrm{SD}}$ of the maximum-likelihood estimator applied to the analog readout outcomes is equal to its asymptotic variance, which saturates the Cram\'er-Rao bound~\cite{cramer1946}:
\begin{align}
	\zeta_{\textrm{SD}} \sim \frac{1}{N F(s_0)}, \label{eq:cramerRaoBound}
\end{align}
where $F(s_0)$ is the Fisher information of the distribution \eqref{eq:distributionS0}:
\begin{align}
	F(s_0) = \statAverage{ \left(\frac{\partial \ln P(\mathcal{O}|s_0)}{\partial s_0}\right)^2 } = - \statAverage{\frac{\partial^2 \ln{P(\mathcal{O}|s_0)}}{\partial s_0^2}}. \label{eq:Fisher}
\end{align}
The last equality in Eq.~\eqref{eq:Fisher} is obtained through integration by parts. Differentiating Eq.~\eqref{eq:distributionS0} twice gives an explicit form for $F(s_0)$:
\begin{align}
	F(s_0) = \frac{1}{4}\int d\mathcal{O}\frac{\left[P(\mathcal{O}|+)-P(\mathcal{O}|-)\right]^2}{P(\mathcal{O}|s_0)}.
\end{align}
Expanding the integrand, we have:
\begin{align}
	F(s_0) = \frac{1}{4}\left[ \int d\mathcal{O} \frac{P(\mathcal{O}|+)^2}{P(\mathcal{O}|s_0)} + \int d\mathcal{O} \frac{P(\mathcal{O}|-)^2}{P(\mathcal{O}|s_0)} - 2\int d\mathcal{O} \frac{P(\mathcal{O}|+) P(\mathcal{O}|-)}{P(\mathcal{O}|s_0)} \right]. \label{eq:fisherInformationExpanded}
\end{align}
When the readout distributions $P(\mathcal{O}|\pm)$ are very well-separated, the Fisher information only contains the shot noise contribution $F(s_0) = 1/(1-s_0^2)$. We isolate this contribution in Eq.~\eqref{eq:fisherInformationExpanded} and upon simplification we find:
\begin{align}
	F(s_0) = \frac{1}{1-s_0^2}-\frac{1}{1-s_0^2} I ,\;\;\;\; I= \int d\mathcal{O} \frac{P(\mathcal{O}|+)P(\mathcal{O}|-)}{P(\mathcal{O}|s_0)},
\end{align}
where $I$ is an overlap integral containing all information about the intrinsic measurement noise described by $P(\mathcal{O}|\pm)$. Therefore, the asymptotic mean squared error of the maximum-likelihood estimator applied to the analog readout outcomes is:
\begin{align}
	 \zeta_{\textrm{SD}} \sim \frac{1-s_0^2}{1-I}. \label{eq:softDecodedMSE}
\end{align}

\subsection{Bias-corrected soft average}
Another possible estimator for the qubit expectation value is the soft average discussed in Ref.~\cite{ryan2013}:
\begin{align}
	s_{\textrm{SA}} = \frac{1}{N}\sum_{i=1}^{N} \mathcal{O}_i. \label{eq:softAverageBiased}
\end{align}
We compare the performance of this estimator to the previously discussed estimators, $s_{\textrm{TA}}$ and $s_{\textrm{SD}}$, for completeness. The expectation value of Eq.~\eqref{eq:softAverageBiased} with respect to $P(\mathcal{O}|s_0)$ has the form:
\begin{align}
	\statAverage{s_{\textrm{SA}} } = A s_0 + B,
\end{align}
where:
\begin{align}
	A = \frac{{\statAverage{ \mathcal{O} }}_+ - { \statAverage{ \mathcal{O} }}_-}{2},\;\;B = \frac{{\statAverage{ \mathcal{O} }}_+ + {\statAverage{ \mathcal{O} } }_-}{2}. \label{eq:scalingCoefficients}
\end{align}
Here, we define the conditional expectations $\statAverage{\mathcal{O}}_{\pm} = \int d\mathcal{O}\,P(\mathcal{O}|\pm)\mathcal{O}$. Thus, the soft average of Eq.~\eqref{eq:softAverageBiased} is biased for general readout probability distributions $P(\mathcal{O}|\pm)$.

To obtain an unbiased estimate, we replace Eq.~\eqref{eq:softAverageBiased} by the soft average of the rescaled values $\mathcal{O}_i'=(\mathcal{O}_i-B)/A$:
\begin{align}
	s_{\textrm{SA}}=\frac{1}{N}\sum_{i=1}^{N} \mathcal{O}_i' = \frac{1}{N}\sum_{i=1}^{N} \frac{\mathcal{O}_i-B}{A}. \label{eq:softAverageUnbiased}
\end{align}
The asymptotic mean squared error $\zeta_{\textrm{SA}}$ of the unbiased soft average estimate, Eq.~\eqref{eq:softAverageUnbiased}, is equal to its asymptotic variance and is given by the central limit theorem:
\begin{align}
	\zeta_{\textrm{SA}}=\statAverage{ {\Delta s}_{\textrm{SA}}^{2} } = \frac{\Delta \mathcal{O}'^2}{N} = \frac{\statAverage{ \mathcal{O}'^2 } - s_0^2}{N}.
\end{align}
In terms of the original observable $\mathcal{O}$, this becomes:
\begin{align}
	\zeta_{\textrm{SA}} = \frac{\Delta \mathcal{O}^2}{A^2 N} = \frac{\statAverage{\mathcal{O}^2 }-(A s_0 + B)^2}{A^2 N}. \label{eq:softAverageMSE}
\end{align}
In the special case of the Gaussian readout, Eq.~\eqref{eq:gaussianDistributions}, we have $A=1$ and $B=0$. Direct calculation of $\statAverage{\mathcal{O}^2}$ then yields the result given in Ref.~\cite{ryan2013}:
\begin{align}
	\zeta_{\textrm{SA}} = \frac{1 + r^{-1} - s_0^2}{N}.
\end{align}
\begin{suppfigure}
\centering
\includegraphics[width=0.6\columnwidth]{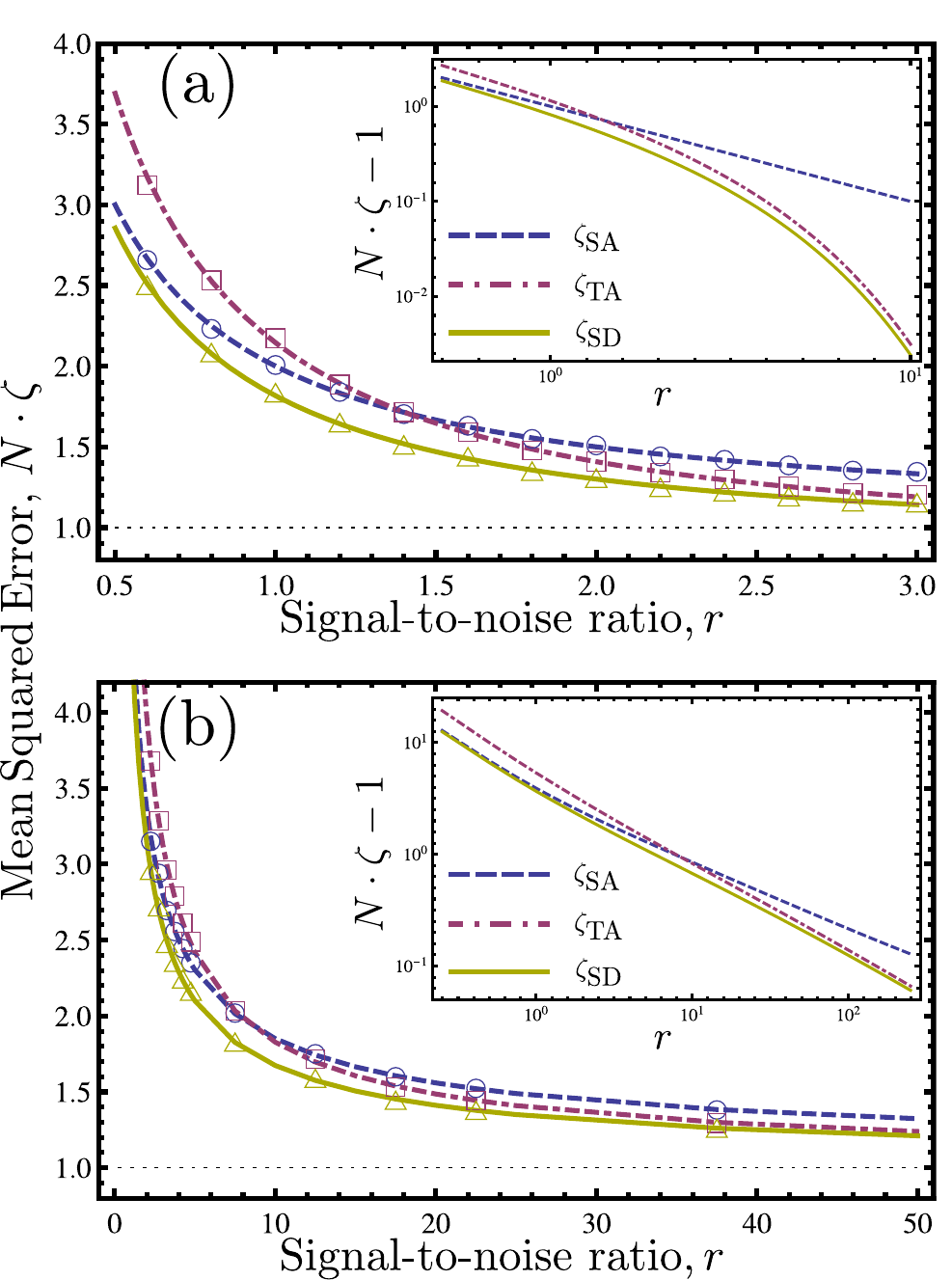}
\centering
\caption{(Color online) Comparison of the asymptotic mean squared error of the soft average $s_{\textrm{SA}}$ (dashed blue), Eq.~\eqref{eq:softAverageMSE}, to that of the maximum-likelihood estimates $s_{\textrm{TA}}$ (dot-dashed magenta), Eq.~\eqref{eq:thresholdedMSE}, and $s_{\textrm{SD}}$ (solid gold), Eq.~\eqref{eq:softDecodedMSE}, for (a) the Gaussian readout and (b) the peak-signal readout. The finite-$N$ MSEs for the soft average $s_{\textrm{SA}}$ (blue circles), the thresholded average $s_{\textrm{TA}}$ (magenta squares) and the soft-decoded estimate $s_{\textrm{SD}}$ (gold triangles) are obtained from $5\times 10^4$ randomly generated measurement records with $N=100$. Inset: Asymptotic MSEs on a logarithmic scale. \label{fig:figS6}}
\end{suppfigure}
Fig.~\ref{fig:figS6} compares the asymptotic performance of the soft average $s_{\textrm{SA}}$ to that of the maximum-likelihood estimates $s_{\textrm{TA}}$ and $s_{\textrm{SD}}$ as a function of the signal-to-noise ratio $r$, for both the Gaussian and the peak-signal readouts. As noted in Ref.~\cite{ryan2013}, the soft average outperforms the thresholded average $s_{\textrm{TA}}$ for low $r$. This is because the distribution $P(\mathcal{O}|s_0)$ approaches a Gaussian centered at $s_0$ when $r\rightarrow 0$ for both readouts, $P(\mathcal{O}|s_0) \simeq \sqrt{\frac{r}{2\pi}} e^{-\frac{(\mathcal{O}-s_0)^2 r}{2}}$, and the maximum-likelihood estimator for the mean of a Gaussian coincides with the soft average. In that case, the soft average $s_{\textrm{SA}}$ is therefore the same as the soft-decoded estimate $s_{\textrm{SD}}$. However, the soft average estimate offers suboptimal performance for finite $r$ and suffers from an significant loss in performance compared to $s_{\textrm{TA}}$ and $s_{\textrm{SD}}$ when $r$ becomes large. In contrast, the soft-decoded estimate $s_{\textrm{SD}}$ is optimal for all $r$.

\clearpage

\end{document}